\documentclass[fp,twocolumn]{jpsj3}
\usepackage{txfonts}

\usepackage{graphicx,color}
\usepackage{bm}
\usepackage{braket}

\title{Chiral Magnetic Effect due to Inhomogeneous Magnetic Fields in Noncentrosymmetric Weyl Semimetals}
\author{Yohei Ibe\thanks{ibe@scphys.kyoto-u.ac.jp} and Hiroaki Sumiyoshi}
\inst{Department of Physics, Kyoto University, Kyoto 606-8502, Japan}

\date{\today}

\abst{
The chiral magnetic effect is a phenomenon where an electromagnetic current is generated along a magnetic field.
 Recently, in nonequilibrium systems, negative longitudinal magnetoresistance has been observed experimentally in Dirac/Weyl semimetals, which provides evidence for the chiral magnetic effect as a nonequilibrium current.
On the other hand, the emergence of the chiral magnetic effect as an equilibrium current is still controversial. We propose a possible realization of the chiral magnetic effect as an equilibrium current using inhomogeneous magnetic fields.
By employing tight-binding calculations and linear response theory, we demonstrate that a finite current density is generated by inhomogeneous magnetic fields, while the spatial integration of the current is equal to zero, which is consistent with the so-called ``no-go theorem'' of the chiral magnetic effect in real lattice systems. Moreover, we propose an experimental setup to detect the effect in Weyl semimetal materials.
}

\begin{document}
\maketitle

\section{\label{sec:level1}Introduction}

Recently, Weyl semimetals (WSMs) have attracted significant interest as  gapless three-dimensional (3D) systems with nontrivial topology.\cite{Murakami2007,Wan2011,Hosur2013,Volovik2003}
WSMs are sometimes referred to as ``a 3D analogue of graphene,'' in contrast to 3D topological insulators, in that WSMs possess gapless linear dispersion not at the surface but in the bulk.
The band structure of WSMs includes nondegenerate Weyl cones and the band-touching points of the cones are called Weyl points (WPs). One can define the chirality for each WP, which plays the role of a topological invariant and takes the value of $1$ or $-1$. It was shown by Nielsen and Ninomiya\cite{Nielsen1983} that  the net chirality in the entire Brillouin zone (BZ) must be zero because of the periodicity of the BZ. Therefore, WPs always appear in pairs with opposite chiralities. 
WPs are ``topologically protected'' in the sense that  Weyl cones do not open the gap in the presence of  weak disorder or interaction unless the WPs with opposite chirality move, come together in the BZ, and eventually pair-annihilate.
After the theoretical proposal that pyrochlore iridates are WSMs by Wan et al.,\cite{Wan2011} a number of candidate WSMs were proposed, some of which have been confirmed experimentally.\cite{Burkov2011a,Kim2013,Xu2011, Jian2015,Sun2015,Wang2016a,Soluyanov2015,Chang2015a,Belopolski2015,Borisenko2015,Huang2016, Singh2012,Wang2016,Ruan2016,Sun2015a,Weng2015,Xu2015s, Lv2015, Yang2015a, Xu2015b, Xu2015c, Xu2015a, Xu2015, Souma2016}
 In particular, TaAs and  similar compounds have recently been confirmed as inversion-symmetry-broken WSMs by angle-resolved photoemission spectroscopy  measurements.\cite{Xu2015s, Lv2015, Yang2015a, Xu2015b, Xu2015c, Xu2015a, Xu2015, Souma2016}
WSMs are expected to have novel transport phenomena, such as an anomalous Hall effect,  \cite{Chen2013,Zyuzin2012,Burkov2011a} chiral magnetic effect (CME),  \cite{Fukushima2008,Zyuzin2012,Zyuzin2012a,Goswami2013,Son2013} and negative longitudinal magnetoresistance.\cite{Son2013}  Among these effects, we deal with the CME in this work.

The CME is a phenomenon where an electrical current is generated along applied magnetic fields and originates from the chiral anomaly in quantum field theory.\cite{Fukushima2008,Zyuzin2012} While the effect was originally discussed  in the context of nuclear physics, it was proposed recently that the CME can occur in noncentrosymmetric WSMs by Zyuzin and cowerkers\cite{Zyuzin2012,Zyuzin2012a} in the framework of linearized continuum theory.
According to their discussion, the relation between the applied magnetic field and the current induced via the CME is given by 
$\bm{j}=-e^2b_0 \bm{B}/2\pi^2 \hbar^2,$
where $b_0$ is the energy splitting of each WP.
This equation is, however,  only valid in the framework of low-energy effective theory. It has been proven by subsequent research that, in real lattice systems, the ground-state current vanishes identically, owing to the periodicity of the BZ.\cite{Vazifeh2013,Yamamoto2015} This is the so-called ``no-go theorem" of the CME.

One should note, however, that if one considers nonequilibrium situations, this no-go theorem is no longer applicable. There have been some recent attempts to realize the CME via dynamical approaches.\cite{Goswami2015,Sekine2016,Taguchi2016, Ebihara2016,Fujita, Xiong2015, Li2016}  Among them, in Refs. \citen{Xiong2015} and \citen{Li2016} a negative longitudinal magnetoresistance was observed experimentally, which provides evidence for the occurence of the CME in such systems, 
 since the resistivity is expected to be decreased in the presence of the CME current.

On the other hand, the CME was originally proposed as an equilibrium current without dissipation.\cite{Fukushima2008, Zyuzin2012a} In this paper, we propose a possible realization of such an effect at equilibrium  by means of spatially inhomogeneous magnetic fields.
 As one of us pointed out in Ref. \citen{Sumiyoshi2016},  in  the no-go theorem of the CME,  only the generation of the {\it total} current (i.e., the net current flowing through a certain cross section) is prohibited, while that of the {\it local} current density is not. Apparently, under uniform magnetic fields, the absence of a net current is equivalent to that of a local current density. Thus, we consider the case where the external magnetic field is spatially nonuniform.

 In this paper, we investigate the electromagnetic response of WSMs under inhomogeneous magnetic fields  via two approaches: a tight-binding model calculation and linear response theory.  In the tight-binding calculation, we numerically calculate the current density of each lattice site under inhomogeneous magnetic fields, which results in a finite current density. Meanwhile, the net current vanishes as a whole, which is consistent with the no-go theorem of the CME. Moreover, in the linear response calculation, we confirm that the CME coefficient $\alpha$, which connects the applied magnetic field and the current density as $\bm{j}=-\alpha \bm{B}$, has a finite value only when the magnetic field is nonuniform. Hence, it follows that  inhomogeneous magnetic fields clearly realize the CME as the ground-state current in WSMs.

The remainder of this paper is organized as follows. In Sect. \ref{sec:lattice}, we explain the framework of the lattice calculation of a four-band model of WSMs under inhomogeneous magnetic fields and the results are shown. The linear response calculation in the same situation is given in Sect. \ref{sec:Kubo}. In Sect. \ref{sec:experiment}, we present an experimental setup to observe the CME as the ground-state current in WSMs. Section \ref{sec:discussion} is devoted to conclusions.

\section{\label{sec:lattice} Lattice Calculation on a Four-Band Model of WSMs}
\subsection{Model and method}
The four-band Hamiltonian of a WSM with a pair of WPs in momentum space is given by 
\begin{eqnarray}\label{TbHamUt}
\begin{split}
\mathcal{H}&=\sum_{\bm{k}} H(\bm{k})c^\dag _{\bm{k}}c_{\bm{k}}\\
H(\bm{k})&=  2\sigma_z \bm{s}\cdot
\left(\begin{array}{c}\lambda \sin{k_x}\\ \lambda \sin{k_y}\\ \lambda_z \sin{k_z} \end{array}\right)
 + \sigma_x M_{\bm{k}}\\
&\quad + b_0 \sigma_z+\bm{b}\cdot \bm{s}\\
M_{\bm{k}}&= 2t\sum_{\alpha=x,y,z}(1-\cos{k_\alpha}),
\end{split}
\end{eqnarray}
which originates from  the 3D topological insulator Bi$_2$Se$_3$ family.\cite{Vazifeh2013} Here, $\bm{\sigma}$ and $\bm{s}$ are the Pauli matrices in orbital and spin space, and $b_0$ and $\bm{b}$  denote the splitting of WPs in energy and momentum respectively, and we assume that the lattice constant is 1.
The Taylor expansion of this Hamiltonian in the vicinity of the WPs gives
\[
H(\bm{k}\approx 0)=\sigma_z \bm{s}\cdot \bm{k} + b_0 \sigma_z+\bm{b}\cdot \bm{s} ,
\]
which coincides with the model of WSMs originating from the multilayer heterostructure of the topological insulator, proposed by Zyuzin and Burkov.\cite{Zyuzin2012} Here, the  Fermi velocity is absorbed in the definition of the wavenumber $\bm{k}$.

We consider the case where the time-reversal symmetry (TRS) is preserved while the inversion symmetry (IS) is not, namely, $\bm{b}=0$ but $b_0 \neq 0$. This corresponds to the situation where a pair of WPs with energy $\pm b_0$ and opposite chiralities exists at the $\Gamma$ point in the BZ.

\begin{figure}[tbp]
  \begin{center}
          \includegraphics[clip, width=7.5cm]{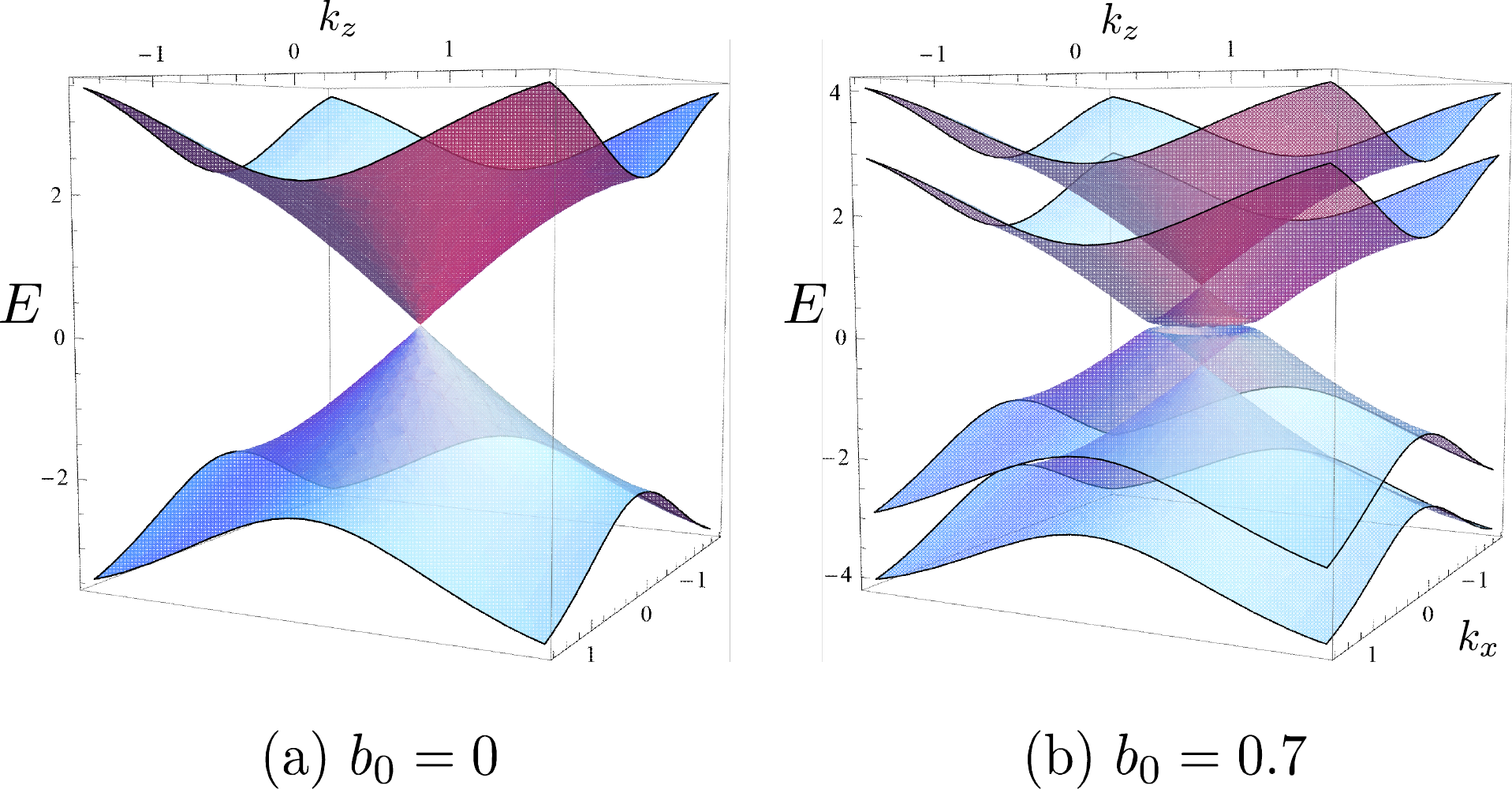}\label{0band}
    \caption{(Color Online) Bulk band dispersion around the $\Gamma$ point obtained from the Hamiltonian (\ref{TbHamUt}) for (a) $b_0=0$ and  (b) $b_0=0.7$. Here,  $k_y$ is fixed as 0 and only the $k_x$- and $k_z$-dependences are shown. }
    \label{bulk}
  \end{center}
\end{figure}

The bulk band spectrum obtained from the Hamiltonian in Eq. (\ref{TbHamUt}) is shown in Fig. \ref{bulk}. Note that if $b_0=0$, i.e., both the TRS and IS are preserved [see Fig. \ref{bulk}(a)], all bands are doubly degenerate owing to the Kramers theorem, and a pair of WPs with the same energy $E=0$ exists at the $\Gamma$ point. On the other hand, if the IS is broken [see Fig. \ref{bulk}(b)], i.e., $b_0\neq0$, the Kramers degeneracy is lifted except at the $\Gamma$ point, and the two WPs have different energies $\pm b_0$. Strictly speaking, in this model,  there is an energy gap at $E=0$ and actually the system is insulating if $b_0\neq 0$. The magnitude of this gap increases with $b_0$. However, this fact does not affect our final result since WPs still exist in the BZ.


Let us consider the case where an external magnetic field is applied in the $z$-direction, which spatially varies along the $x$- or $y$-direction. In such a case, in general, the translational symmetries in the $x$- and $y$-directions are broken. 
Thus, we perform a Fourier transformation of the Hamiltonian to obtain a real-space representation in the $x$- and $y$-directions, which gives
\begin{equation}
\begin{split}
\mathcal{H} &=\sum_{x,y,k_z} \Bigl[ \{2t  (3 -\cos{k_z})\sigma_x +2\lambda_z \sin{k_z}\sigma_z s_z \\
&\qquad +b_0 \sigma_z\}c^\dag_{x,y,k_z}c_{x,y,k_z}\\
&\qquad+\{(-t\sigma_x -i\lambda \sigma_z s_x)c^\dag_{x+1,y,k_z}c_{x,y,k_z}+{\rm h.c.}\}\\
&\qquad+\{(-t\sigma_x -i\lambda \sigma_z s_y)c^\dag_{x,y+1,k_z}c_{x,y,k_z}+{\rm h.c.}\}\Bigr].
\label{finite_ham}
\end{split}
\end{equation}
The effects of external magnetic fields are introduced via the conventional Peierls substitution for the hopping matrix elements,
\[
t_{\rm hop} \to t_{\rm hop} \exp \left[ \frac{2\pi i}{\Phi_0}\int_i ^j \bm{A} \cdot d\bm{l}\right],
\]
where $\Phi_0=h/e$ stands for the flux quantum and the the integral is taken along the straight line between lattice sites $\bm{r}_i$ and $\bm{r}_j$.

From the Hamiltonian in Eq. (\ref{finite_ham}), the current operator along the $z$-direction is given by
\begin{equation}
\begin{split}
\hat{J_z} &=-e \frac{\partial \mathcal{H}}{\partial k_z}\\
&=\sum_{x, y, k_z}(-e) [2t\sin{k_z} \sigma_x + 2\lambda_z \cos{k_z} \sigma_z s_z]c^\dag_{x,y,k_z}c_{x,y,k_z}\\
&\equiv\sum_{x,y,k_z}\hat{j_z}(x,y,k_z),
\end{split}
\end{equation}
where the operator of the current density along the $z$-direction at each $(x,y,k_z)$ is given by
\begin{eqnarray*}
\hat{j_z}(x, y, k_z)&=&-e [2t\sin{k_z} \sigma_x + 2\lambda_z \cos{k_z} \sigma_z s_z]c^\dag_{x,y,k_z}c_{x,y,k_z}.
\end{eqnarray*}
Therefore, the expectation value of the current density at each point $(x,y)$ is given by
\[
j_z (x,y)=  \sum_{n:{\rm band},kz} \Braket{n,k_z |  \hat{j_z}(x,y,k_z) |n,k_z}n_F[\epsilon_n(k_z)],
\]
where $\ket{n,k_z}$ is the eigenstate of the $n$th band with wavenumber $k_z$, and  $n_F$ is the Fermi distribution function. Here, we calculate the current density only at zero temperature.

\subsection{Inhomogeneous magnetic fields and local current density}\label{tbinhomo}
\begin{figure}[htbp]
\begin{center}
\includegraphics[width=8cm]{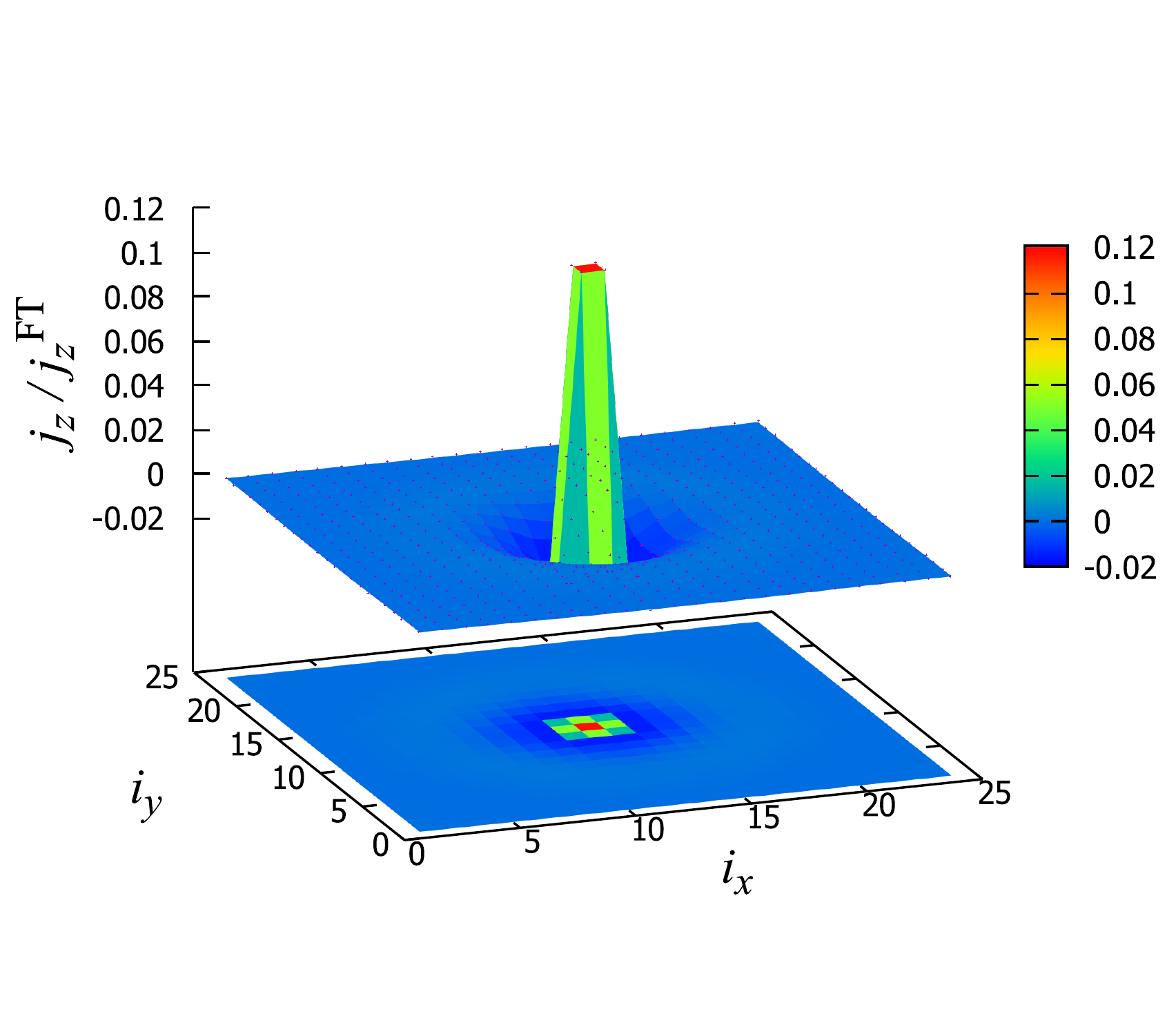}
\end{center}
\caption{(Color Online) Plot of current density under delta-function-shaped magnetic field. $i_x$ and $i_y$ are the site indices along the $x$- and $y$-directions, respectively. A periodic boundary condition is taken in the $z$-direction. The system size is $24 \times 24 \times 100$, and the parameters are set as  $t=0.5$, $\lambda=\lambda_z=1$, $b_0=0.7$. The number of magnetic flux in the unit of flux quantum through the plaquette at the center of the system is 0.1. Here, $j_z^{\rm FT}=-e^2b_0 B_z/(2\pi^2 \hbar^2)$ is the value of $j_z$ predicted from low-energy effective field theory.}
\label{delta}
\end{figure}

We have examined three patterns of the magnetic field configuration. 
First, we apply a magnetic flux tube, which pierces the center of the system along the $z$-direction as follows:
\begin{eqnarray*}
B_z(x,y)&=&B \delta(x) \delta(y).  
\end{eqnarray*}
Figure \ref{delta} shows the spatial property of the calculated current density.
It can be seen that, in the vicinity of the flux, a nonzero local current density  is generated, which flows parallel to the magnetic field.
On the other hand, slightly away from the flux, the current flows in the opposite direction. Further away from the flux, the current density approaches zero.
Here, as we expected, the total current $J_z$ is zero within our numerical precision, which is consistent with the no-go theorem of the CME.

\begin{figure}[t]
\begin{center}
\includegraphics[width=7cm]{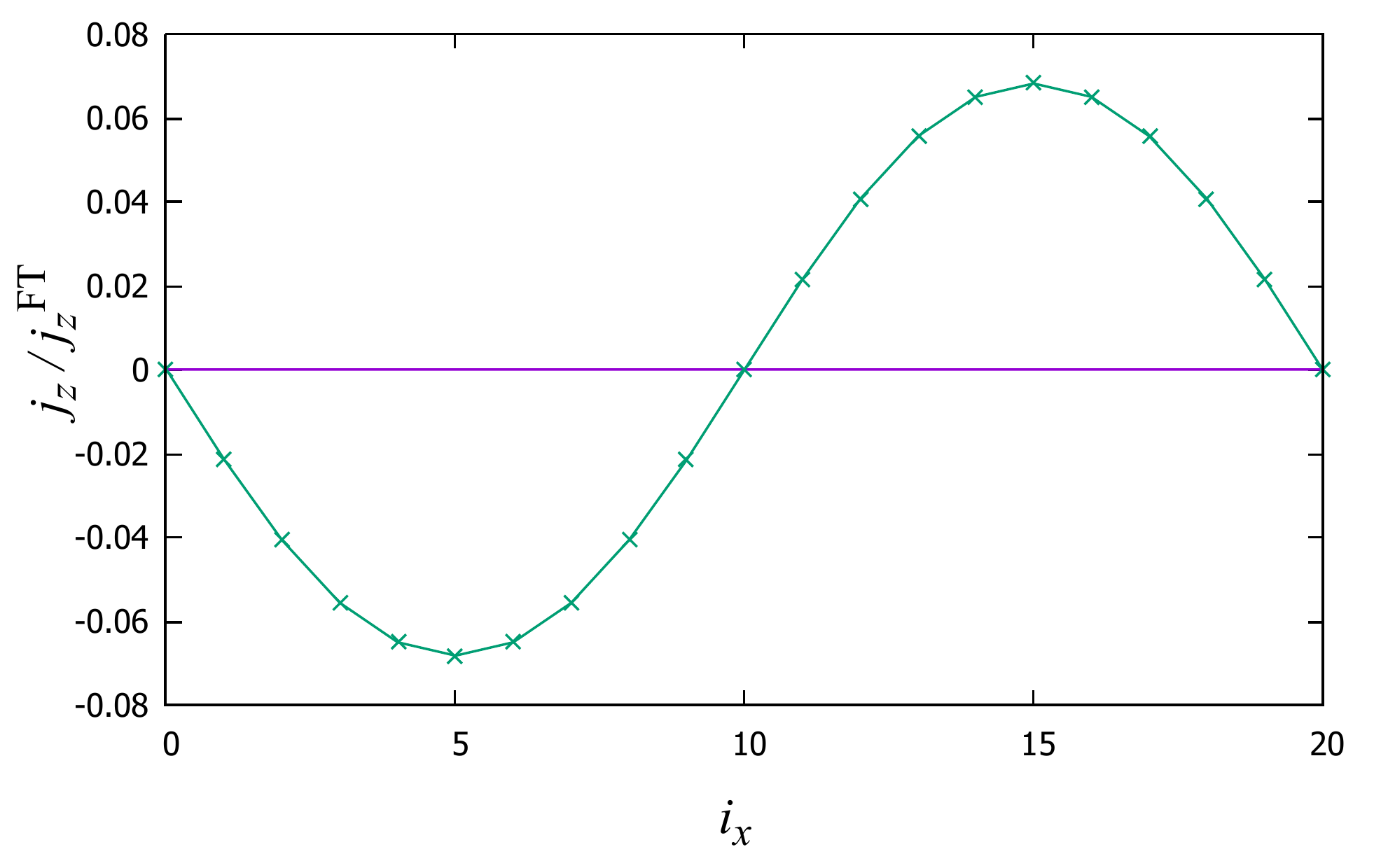}
\end{center}
\caption{(Color Online) Plot of current density under a sinusoidal magnetic field, which varies sinusoidally along the $x$-direction and is constant along the other directions. Here, a periodic boundary condition is taken in the $x$-, $y$-, and $z$-directions. The system size is  $20 \times 10 \times 100$ and the parameters are set as $t=0.5$, $\lambda=\lambda_z=1$, $b_0=0.7$. The maximum value of the flux per plaquette is 0.001.}
\label{sin}
\end{figure}

In Fig. \ref{sin}, we show the results for a magnetic field that varies sinusoidally in the $x$-direction as
\begin{eqnarray*}
B_z(x,y)&=&B\sin{(2\pi x/L_x)},
\end{eqnarray*}
where $L_x$ is the length of the system in the $x$-direction; thus, the phase of the sine function is changed by $2\pi$ from one end to another in the system.
Here, since the magnetic field [and also $j_z(x,y)$] is constant along the $y$-direction, only the $x$-dependence of $j_z(x,y)$ is shown. Similar to the former case, a finite current density is generated but the net current vanishes. 

\begin{figure}[t]
\begin{center}
\includegraphics[width=7cm]{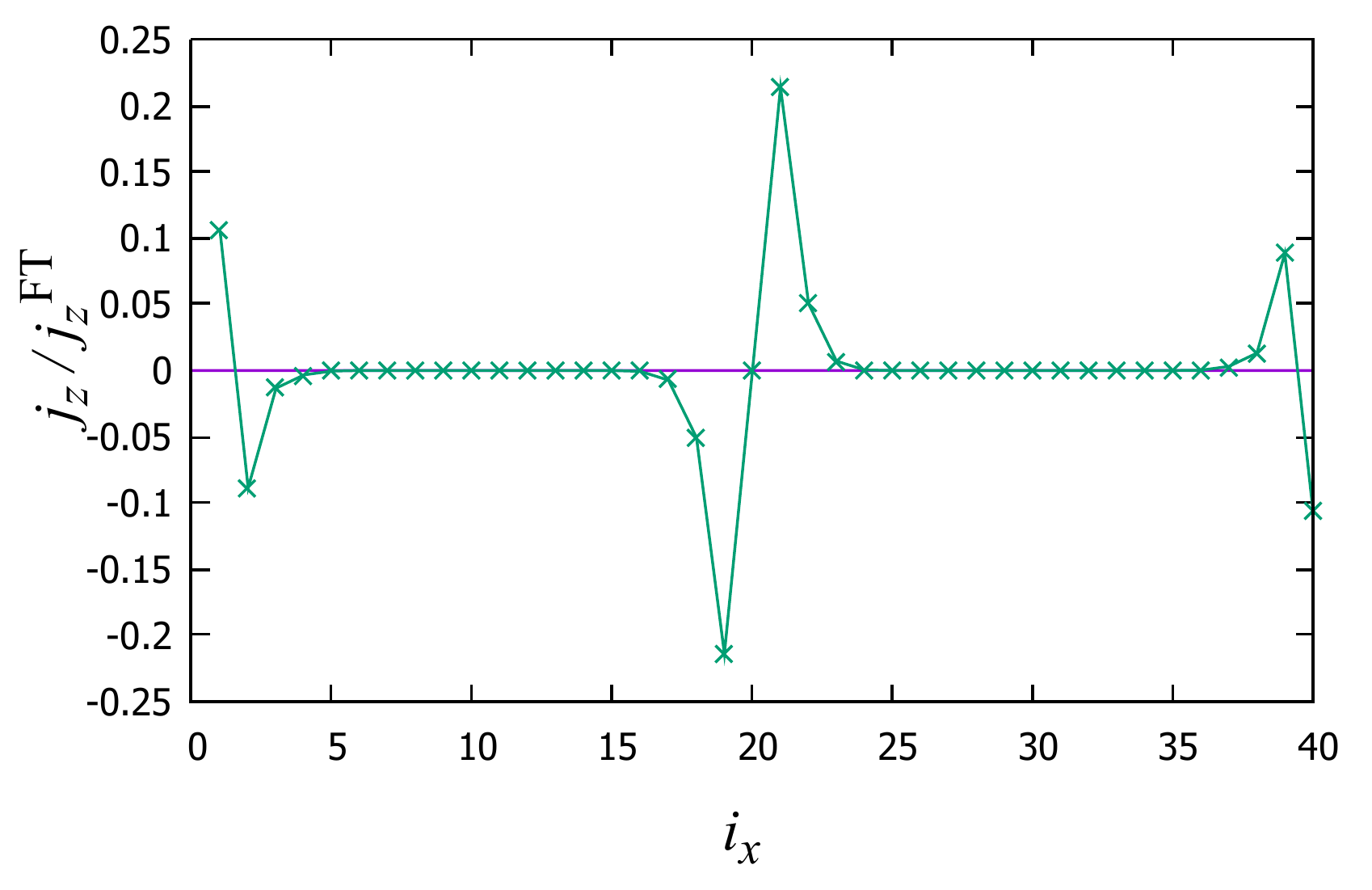}
\end{center}
\caption{(Color Online) Plot of current density under a domain-wall-shaped magnetic field.  Here, we show only the $i_x$-dependence of the current density since it is constant in the other directions. The system size is $40 \times 10 \times 100$, the parameters are set as $t=0.5$, $\lambda=\lambda_z=1$, $b_0=0.7$, and the flux per plaquette is 0.1.}
\label{domain}
\end{figure}

\begin{figure}[t]
  \begin{center}
	\includegraphics[clip, width=8cm]{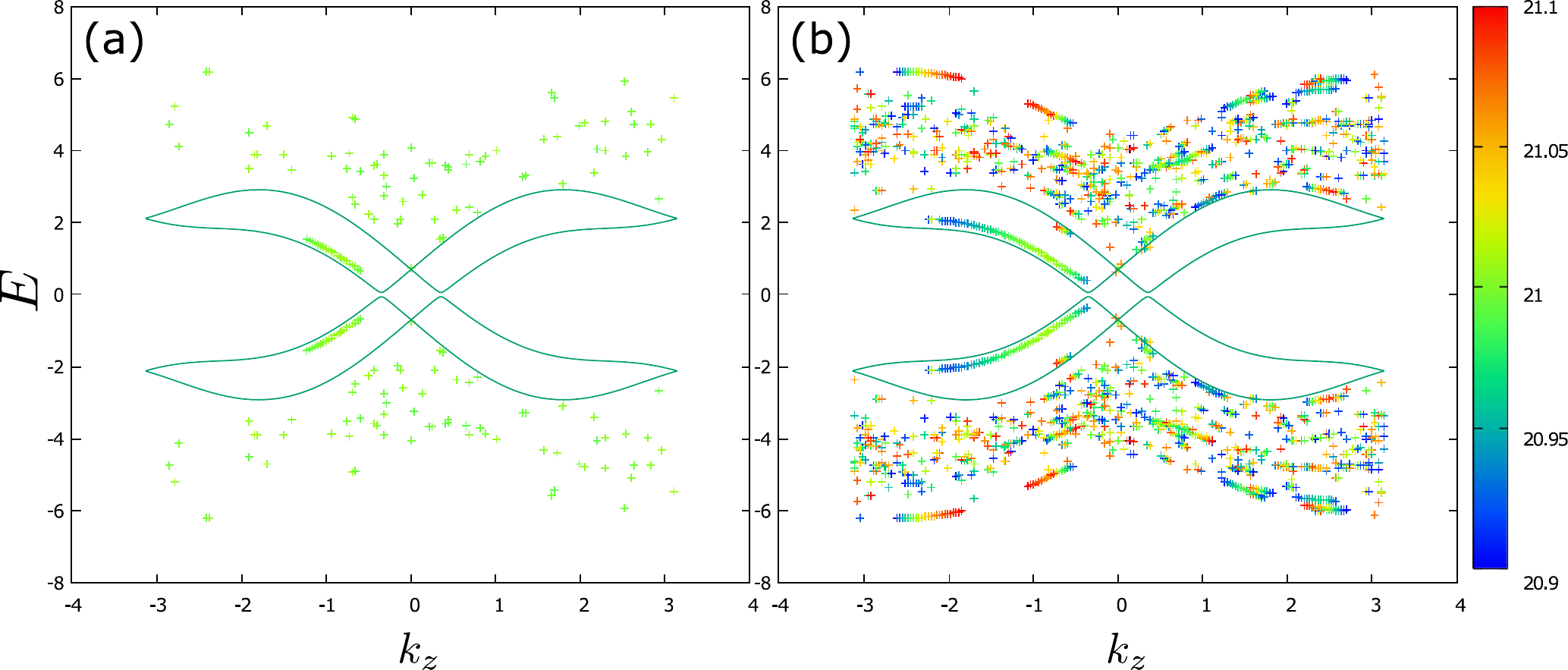}
    \caption{(Color Online) Energy dispersion obtained for a domain-wall-shaped magnetic field. Here, the green curves are the envelope of the bulk spectrum. Since the current density flows in the $+z$-direction around $i_x = 21$, we plot the energy eigenvalues for the eigenstates whose expectation value of the $x$-coordinate $\langle i_x \rangle $ is in the region of (a)  $20.99<\langle i_x \rangle <21.01$ and  (b) $20.9<\langle i_x \rangle <21.1$. The color of each point indicates $\langle i_x \rangle$.}
    \label{band_half}
  \end{center}
\end{figure}

Finally, we divide the system into two parts and apply uniform magnetic fields in opposite directions to each sector.
\begin{eqnarray*}
B_z(x,y)&=&\left\{ \begin{array}{ll}
B & (x>0) \\
-B & (x<0)\\
\end{array} \right.
\end{eqnarray*}
Figure \ref{domain} shows the results for such a magnetic field.
Here, the sign of the magnetic field changes at $i_x=20$. Also in this case, a finite current density is generated while the net current vanishes.  One can observe that the current flows upward (downward) around $i_x= 21$ ($i_x = 19$). To more closely examine  the wavefunctions at $i_x\approx 21$, we plot the energy eigenvalues for the eigenstates,  whose expectation value of the $x$-coordinate  $\langle i_x \rangle \approx 21$  [see Fig. \ref{band_half}]. Here, the color of each point indicates the value of $\langle i_x \rangle$ and  the green lines are the bulk spectrum. One can see that at $\langle i_x \rangle \approx 21$, some linear dispersions appear along the bulk spectrum, which can be identified as chiral modes, i.e., $n=0$ Landau levels.  If we focus only on the filled states with $E<0$, there is a chiral mode with a positive slope, and this mode can be regarded as giving rise to the current in the $z$-direction.

In all the above cases, a finite current density is generated, while the net current is equal to 0, which is in agreement with the no-go theorem. 
We have tested some other configurations of  inhomogeneous magnetic fields and obtained similar results. Note that the magnitude of the calculated current density $j_z$ is comparable to the values  predicted from effective field theory.
Hence, it follows that the spatial variation of the magnetic field is essential for the CME.

\section{\label{sec:Kubo}  Linear Response Calculation}
To examine the magnetic field response of noncentrosymmetric WSMs in a different way, we carried out a linear response calculation with the Kubo formula. 

\subsection{Formulation and method}
We discuss the current response of the WSM to external magnetic fields via the CME following the discussion of Chang and Yang. \cite{Chang2015}

Within the framework of linear response theory, the relationship between the current density and the vector potential that generates an external magnetic field can be represented as
\begin{equation}
j^i (\bm{q},\omega)=\Pi_{i,j}(\bm{q},\omega) A^j(\bm{q},\omega).
\label{jpia}
\end{equation}
Here, Einstein's summation convention is adopted for repeated indices.
$\Pi_{i,j}(\bm{q},\omega)$ is the {\it current-current correlation function}, which can be calculated as
\begin{equation}
\begin{split}
\Pi_{i,j}(\bm{q},i \nu_m) &=\frac{1}{\beta V} \sum_{\bm{k},n} {\rm tr}[\hat{j}_i(\bm{k})  G(\bm{k}+\bm{q}, i \omega_n + i \nu _m) \\
& \quad \times \hat{j}_j (\bm{k}) G(\bm{k}, i \omega_n)],
\end{split}
\label{cccfunc}
\end{equation}
where $V$ is the volume of the system, $\beta =1/k_{\rm B}T$ is the inverse temperature, and  $\omega_n$ and $\nu_m$ are the fermionic and bosonic Matsubara frequencies, respectively. 
Here, the current density operator and Matsubara Green function are given by
\begin{eqnarray}
\hat{\bm{j}}(\bm{k})&=&-\frac{e}{\hbar}\frac{\partial H(\bm{k})}{\partial \bm{k}}\label{jop}  \\
G(\bm{k},i\omega_n)&=& (i\omega_n - H(\bm{k}))^{-1}\label{G},
\end{eqnarray}
respectively, and the four-band Hamiltonian $H(\bm{k})$ is given by Eq. (\ref{TbHamUt}).

Now let us consider the CME coefficient $\alpha(\bm{q},\omega)$, which is defined as
\begin{equation}
\bm{j}_{\rm CME}(\bm{q},\omega)=-\alpha(\bm{q},\omega)\bm{B}(\bm{q},\omega).
\label{jalphab}
\end{equation}
Note that because the current density has odd parity and magnetic field has even parity, the CME coefficient $\alpha$ has odd parity.
In the wavenumber space the magnetic field is $\bm{B}(\bm{q},\omega)=i \bm{q}\times \bm{A}(\bm{q},\omega)$, in other words,
\begin{eqnarray*}
B^i(\bm{q},\omega) &=& i \epsilon^{ijk} q_j A^k(\bm{q},\omega) \\
&=& -i\epsilon^{ijk} q_k A^j(\bm{q},\omega).
\end{eqnarray*}
Substituting this into Eq. (\ref{jalphab}), we obtain
\begin{eqnarray}
j^i(\bm{q},\omega) =i \alpha(\bm{q},\omega) \epsilon^{ijk}q_k A^j(\bm{q},\omega).
\end{eqnarray}
Comparing this with Eq. (\ref{jpia}), we thus find that only the antisymmetric part of $\Pi_{i,j}(\bm{q},\omega)$, i.e., $\Pi_{i,j}^{\rm anti}=(\Pi_{i,j} -\Pi_{j,i}) /2$, gives rise to the CME coefficient $\alpha$ as 
\begin{equation}
\Pi^{\rm anti}_{i,j}(\bm{q},\omega) =i \alpha(\bm{q},\omega) \epsilon^{ijk}q_k ,
\end{equation}
which leads to
\begin{equation}\label{alpha}
\alpha(\bm{q},\omega) = -\frac{i}{q_z}\Pi^{\rm anti}_{xy}(\bm{q},\omega).
\end{equation}

To calculate $\Pi^{\rm anti}_{xy}(\bm{q},\omega)$, we substitute Eqs. (\ref{jop}) and (\ref{G}) into Eq. (\ref{cccfunc}), and then carry out the summation over the Matsubara frequency and the analytical continuation $i\nu_m \to \hbar \omega+i\delta$. Here, since we consider only the case of $\omega=0$, we fix $i \nu_m=0$ from the beginning.

According to the previous research, \cite{Chen2013, Chang2015}  since the CME coefficient $\alpha$ has a singularity at $\omega, \bm{q}= 0$, its value depends on the order of the zero limit of the two variables. More specifically, $\alpha$ vanishes in the {\it static} limit ($\omega \to 0$ first), while it becomes nonzero in the {\it uniform} limit ($ \bm{q} \to 0$ first).

In contrast, we consider the case where the wavenumber in the $x$-direction $q_x$ is kept nonzero while the limit of $q_y, q_z, \omega \to 0$ is taken to examine the effect of the nonuniformity of the magnetic field. This corresponds to the situation that the magnetic field in the $z$-direction $B_z$ depends only on the $x$-coordinate, and is independent of the $y$- and $z$-coordinates.  Since $\bm{q}\neq 0$ (i.e., away from the singularity), it does not matter if we fix $\omega=0$ from the beginning.
From Eqs. (\ref{cccfunc}) and (\ref{alpha}), $\alpha(\bm{q},\omega)$ can be calculated as
\begin{eqnarray}
\begin{split}
\alpha(q_x)&\equiv \lim_{q_z \to 0}\alpha(q_x\neq0, q_y=0,q_z,\omega=0) \\
&=\left. -i\frac{\partial \Pi^{\rm anti}_{xy}}{\partial q_z}\right|_{q_z=0} \\
&=-\frac{i}{2\beta V}\sum_{\bm{k},n} {\rm tr}[\hat{j}_x(\bm{k}) \frac{\partial G}{\partial k_z}(\bm{k}+q_x \hat{k}_x, i \omega_n)  \\
&\quad\times  \hat{j}_y (\bm{k}) G(\bm{k}, i \omega_n)-(x \leftrightarrow y)]  \\
&=-\frac{i}{2\beta V}\sum_{\bm{k},n} {\rm tr}[(\hat{j}_x(\bm{k})G(\bm{k},i\omega_n)\hat{j}_y(\bm{k})-(x \leftrightarrow y))   \\
&\quad\times G(\bm{k}+q_x \hat{k}_x, i \omega_n)\hat{j}_z(\bm{k}) G(\bm{k}+q_x \hat{k}_x, i \omega_n)],\label{alphaCME}
\end{split}
\end{eqnarray}
where $\hat{k}_x$ is the unit vector in the $k_x$-direction and  the relation  $\Pi^{\rm anti}_{xy}(q_x\neq 0,q_y=q_z=0,\omega=0)=0$ is used.
We calculate this numerically for the four-band  Hamiltonian in Eq. (\ref{TbHamUt}).

\subsection{Effect of inhomogeneous magnetic fields on the CME coefficient $\alpha$}

\begin{figure}[t]
\begin{center}
\includegraphics[width=8cm]{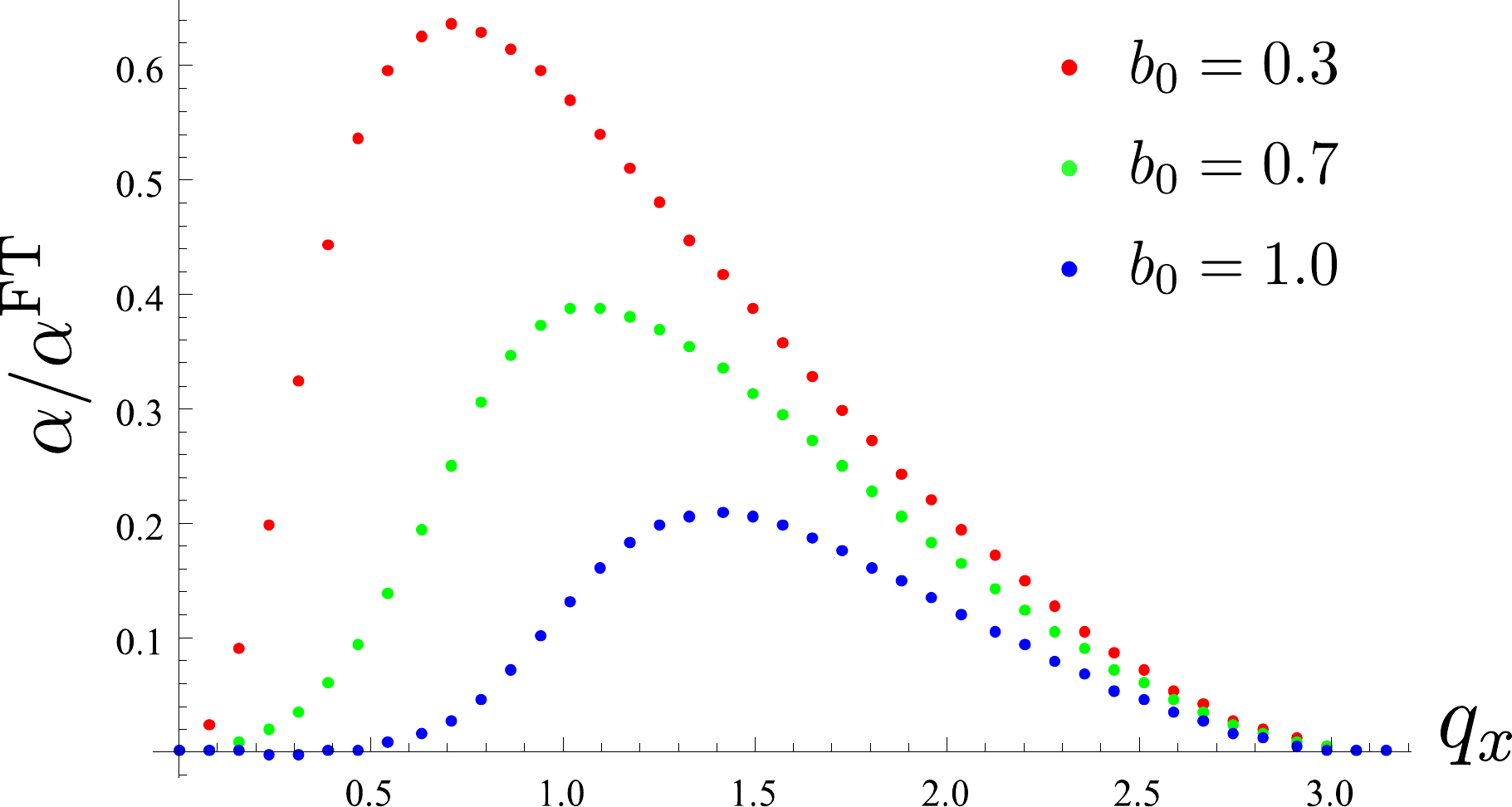}
\end{center}
\caption{(Color Online) $q_x$ dependence of the CME coefficient $\alpha$ in units of $\alpha^{\rm FT}=e^2b_0/(2\pi^2\hbar^2)$. Here, the parameters are set as  $t=0.5$, $\lambda=\lambda_z=1$, $T=0.1$ K.}
\label{4band_alphaqx}
\end{figure}

The $q_x$ dependence of the CME coefficient $\alpha(q_x)$ for different values of $b_0$ is shown in Fig. \ref{4band_alphaqx}. 
It can be seen that $\alpha$ takes a finite value at a finite $q_x$. Meanwhile, $\alpha$ vanishes in the limit $q_x \to 0$, which is consistent  with the no-go theorem of the CME in real lattice systems.

Of course, if either $q_x$ or $b_0$ is equal to zero, $\alpha$ gives 0. In other words, both the IS breaking and the inhomogeneity of applied magnetic fields are essential for the CME.
Moreover, similar to the result of the lattice calculations, the magnitude of the CME coefficient $\alpha$ has comparable values to  $\alpha^{\rm FT}=e^2b_0/(2\pi^2 \hbar^2)$, which was predicted by low-energy effective theory.

Hence, it is confirmed that spatially varying magnetic fields clearly generate finite electrical currents in noncentrosymmetric  WSMs.

\section{\label{sec:experiment}Experimental Implications}
We propose an experimental setup to detect the CME as the ground-state current in WSMs that employs scanning superconducting quantum interference device (SQUID) measurement, which can detect weak inhomogeneous magnetic fields.\cite{Vasyukov2013}
The candidate material is SrSi$_2$, which has WPs with opposite chirality located at different energies owing to the lack of both inversion and mirror symmetries. \cite{Huang2016}
\begin{figure}[t]
\begin{center}
\includegraphics[width=8.7cm]{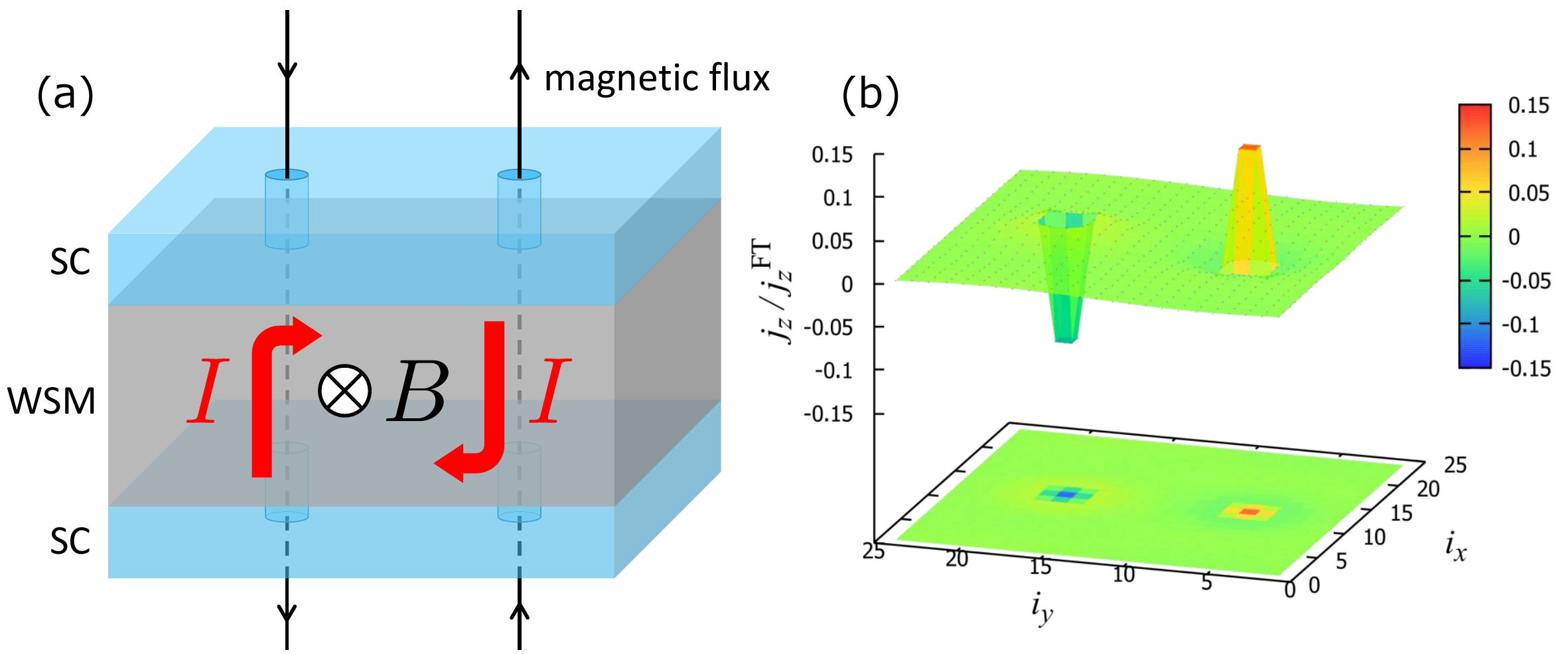}
\caption{(Color Online) (a) Experimental setup to detect the CME due to inhomogeneous magnetic fields via scanning SQUID measurement. Here, $I$ and $B$ are the CME current and the induced magnetic field, respectively. \quad  (b) Tight-binding calculation of current density for an antiparallel pair of delta-function-shaped magnetic fields. The parameters are the same as those in Fig. \ref{delta}. }
\label{setup}
\end{center}
\end{figure}

Our proposal for the experimental setup is shown in Fig. \ref{setup}(a).
We consider a structure where a sample of the WSM SrSi$_2$ is sandwiched between superconductors with two parallel small holes, fabricated by the focused ion beam technique, which can be made as small as 750 nm diameter.\cite{Jang2011}
 If we apply magnetic fields with the same magnitude in opposite directions from each hole, quantized magnetic fluxes will penetrate into the WSM, which generates a pair of local currents.
Figure \ref{setup}(b) shows the result of a tight-binding calculation for such a setup.
Here, the total current due to one flux becomes finite, while the sum of the contributions from both fluxes is zero.
This local current circulates in the finite-size sample, which induces a magnetic field in the direction perpendicular to the plane of the page.
The magnitudes of the circulating current and the induced magnetic field are estimated to be $I\sim 10^{-7}$ A and $B\sim 10^{-8}$ T, respectively. Here we used the material parameter $b_0 \sim 0.1 $ eV and the lattice constant $a \sim 10$ $ \AA$. The hole diameter and the distance between the holes on the superconductor are set as 750 nm and 10 ${\rm \mu}$m, respectively, and we assume that a single flux quantum is introduced in each hole.\cite{Jang2011}   It is feasible to detect $B \sim 10^{-8}$ T via the scanning SQUIDs.

\section{\label{sec:discussion}  Conclusions}

In this paper, we have challenged a common belief regarding the CME:  the absence of a ground-state current in real lattice systems. Employing the tight-binding calculation and linear response theory, we have clarified that if an applied magnetic field is spatially nonuniform, a ground-state current is generated  parallel to the magnetic field, i.e., the CME occurs even at equilibrium. 
Here, the magnitude of the current  is of a comparable order to that predicted from low-energy effective theory. 
It should be stressed that the net current is equal to zero, which is consistent with the no-go theorem of the CME. 
Furthermore, the induced current is experimentally observable in realistic materials via the scanning SQUID measurement.

\section*{Acknowledgements}
We are grateful to A. Shitade, T. Nomoto, H. Fujita, Y. Nagai, Y. Yanase, and  N. Kawakami for helpful discussions. This work was supported by Grants-in-Aid for Scientific Research from MEXT of Japan [Grant No. JP15H05855 (KAKENHI on Innovative Areas ``Topological Materials Science'')]. HS is supported by a
JSPS Fellowship for Young Scientists (14J00647).

\bibliography{refCME}
\bibliographystyle{jpsj}
\end{document}